\documentclass[doublecol,figures,dvipdfm]{epl2} 

\usepackage{amstext}
\usepackage{dcolumn}
\usepackage{bm}
\newcommand{\vs}{{\it vs.}}

\newcommand{\al}{{\it et al.}}
\newcommand{\bq}{\begin{equation}}
\newcommand{\eq}{\end{equation}}

\title{Short-fragment Na-DNA dilute aqueous solutions: fundamental length scales and screening}
\shorttitle{Short-fragment Na-DNA solutions} 

\author{S.~Tomi\'{c}\inst{1}\thanks{E-mail: \email{stomic@ifs.hr}} \and
        S.~Dolanski~Babi\'{c}\inst{1}\thanks{Permanent address: Department of Physics and Biophysics, Medical School, University of Zagreb - Zagreb, Croatia} \and
        T.~Ivek\inst{1} \and
        T.~Vuleti\'{c}\inst{1} \and
        S.~Kr\v{c}a\inst{2} \and
        F.~Livolant\inst{3} \and
        R.~Podgornik\inst{4,5,6}
}

\shortauthor{S. Tomi\'{c} \etal}	

\institute{
  \inst{1} Institut za fiziku - HR-10001 Zagreb, Croatia\\
  \inst{2} Rudjer Bo\v{s}kovi\'{c} Institute - HR-10001 Zagreb, Croatia\\
  \inst{3} Laboratoire de Physique des Solides, Universit\'{e} Paris Sud - F-91405 Orsay, France\\
  \inst{4} Department of Physics, University of Ljubljana - SI-1000 Ljubljana, Slovenia\\
  \inst{5} J. Stefan Institute - SI-1000 Ljubljana, Slovenia\\
  \inst{6} Laboratory of Physical and Structural Biology, NICHD, National Institutes of Health - Bethesda, MD 20892, USA}


\abstract{Dielectric spectroscopy is used to investigate fundamental length
scales of $146\un{bp}$ short-fragment (nucleosomal) dilute Na-DNA solutions.
Two relaxation modes are detected: the high- and the low-frequency mode.
Dependence of the corresponding length scales on the DNA and on the (uni-valent)
salt concentration is studied in detail, being different from the case of long,
genomic DNA, investigated before. In low added salt regime, the length scale of
the high-frequency mode scales as the average separation between DNAs, though it
is smaller in absolute magnitude, whereas the length scale of the low-frequency
mode is equal to the contour length of DNA. These fundamental length scales in
low added salt regime do not depend on whether DNA is in a double stranded or
single stranded form. On the other hand, with increasing added salt, the
characteristic length scale of low-frequency mode diminishes at low DNA
concentrations probably due to dynamical formation of denaturation bubbles
and/or fraying in the vicinity of DNA denaturation threshold.}

\begin{document}

\maketitle

Semiflexible polyelectrolytes are fundamental components of biological
environment, ranging from charged polymers such as deoxyribonucleic acid (DNA)
and proteins, all the way to molecular aggregates such as bacterial fd viruses
and the tobacco mosaic virus~\cite{Daune03}. DNA is in many respects a paradigm
of a semiflexible highly charged polymer. In aqueous solutions it assumes a
conformation of an extended statistical coil, whereas {\it in vivo} quite long
genomic DNA is usually folded in dense and compact states to fit within the
micron-sized nucleus of eukaryotic cells or even smaller viral capsids~\cite{Bloomfield00}.
Such wide range of complex behaviors of DNA is due to its connectivity,
stiffness and strong electrostatic interactions. To a large degree this behavior
seen {\it in vivo} can be closely reproduced {\it in vitro} by tuning the DNA
concentration, varying the amount of added salt, as well as valency of the
counterions~\cite{ Bloomfield00,Dobrynin}. A full description and understanding
of single DNA chain structure together with the structural organization of DNA
chains in aqueous solutions are of fundamental importance in the study of living
systems.

Let us first reiterate the results of a recent dielectric spectroscopy study of
polydisperse Na-DNA~\cite{Tomic06,Tomic07} in the case of {\bf semidilute}
aqueous solutions of long ($2-20\un{kbp}$), genomic DNA. It revealed two
relaxation modes that can be attributed to diffusive motion of DNA counterions,
with fundamental length scales consistent with theoretical estimates~\cite{Dobrynin,Odijk77,deGennes76}. 
Both relaxation modes are found to be strongly DNA concentration-dependent,
setting them apart from previously observed concentration-independent processes
that are contingent on the degree of polymerization $N$ (number of monomers in a
given molecule)~\cite{Takashima66}. The measured fundamental length scale,
probed by the low-frequency (LF) mode, characterizes single chain properties,
being equal to the size of the Gaussian chain composed of correlation blobs that
scales as $c_{\mathrm{DNA}}^{-0.25}$ in the low added salt limit~\cite{Dobrynin}.
In the high added salt limit the LF mode length scale equals the persistence
length $L_{\mathrm{p}}$ and scales as $L_{\mathrm{p}} = L_0 + a I_{\mathrm{s}}^{-1}$ (in \AA) which is nothing
but the well-known Odijk-Skolnick-Fixman (OSF) result. Here $L_0$ is close to
the structural persistence length of $500\un{\mbox{\AA}}$ and $I_{\mathrm{s}}$ is the ionic
strength of the added salt (in M). On the other hand, the high-frequency (HF)
mode is probing the collective properties of the DNA solution which are
characterized by the de Gennes-Pfeuty-Dobrynin (GPD) correlation length or the
mesh size, that scales as $c_{\mathrm{DNA}}^{-0.5}$. At low DNA concentrations
and in the low added salt limit the LF mode length scale reflects the locally
fluctuating DNA regions with partially exposed hydrophobic core that yields the
scaling form $c_{\mathrm{DNA}}^{-0.33}$. Even when the denaturation protocol is
applied, unzipping of the two DNA strands appears to be at most local and
complete separation of the strands in semidilute solutions is never really
accomplished. For both modes the high and the low salt are characterized by the
competition between added salt and intrinsic DNA counterions. 

Due to the size of DNA used in these experiments, all DNA length
dependence was already saturated and none of the two modes show any. For shorter
chains other experiments show that even $N$-dependent modes can often be
concentration-dependent~\cite{Molinari81}, leading to a complicated
interdependence of the two effects. Another intriguing issue is if and how the
screening and fundamental length scales will change in the limit of very low DNA
density, that is, below the semidilute-dilute crossover. Contrary to the
semidilute regime, where polyelectrolyte chains are in general entangled with
each other, the dilute regime with no or low added salt is characterized
by extended DNA conformations where each chain is well separated from all other
ones, leading to an average separation between chains that scales as
$c_{\mathrm{DNA}}^{-0.33}$~\cite{Dobrynin}. In addition, the low concentration
of chains also affects the characteristics of the Manning-Oosawa counterion
condensation which in fact takes place in two separate zones of volume associated
with each chain~\cite{Deshkovski}.

In this Letter we address some of these issues while trying to
characterize the structure of {\bf dilute} DNA solutions composed of short-fragment
DNA. To this effect we used nucleosomal DNA ($\sim 146\un{bp}$) chains prepared
as described previously~\cite{Sikorav94}. The low protein content was verified
and DNA concentration was determined by UV spectrophotometry. We perform a
systematic investigation of how the dielectric properties of these monodisperse
short-fragment DNA aqueous solutions evolve upon change of DNA concentration and
added salt over a range of two to three orders of magnitude. DNA solutions were
prepared as described previously (see Materials and Methods in ref.~\cite{Tomic07},
preparation protocols I and II.3). Since the contour length $L_{\mathrm{c}}$ of DNA
chains is on the order of $500\un{\mbox{\AA}}$, the concentration of
polyelectrolyte solutions that we deal with here is always below the chain
overlap concentration $c^\ast$~\cite{Dobrynin}. A very crude estimate of $c^\ast$
based on de Gennes arguments~\cite{deGennes76} yields $c^\ast$ of the order of
$1\un{mg/mL}$, which is close to the upper concentration bound in our
experiments~\cite{note1}. This means that we are effectively always in dilute
regime which, in contrast to the semidilute one, has not been much studied
experimentally. Concurrently we have also considered to what extent the
interpretation of our results depends on DNA conformation in low salt and in pure
water solutions, more specifically on whether DNA is in a single stranded or
double stranded form. Denaturation studies (performed by dielectric
spectroscopy as described previously~\cite{Tomic07}) indicate that for these
dilute conditions DNA double-helix is denatured in pure water solutions for
DNA concentrations below about $0.4\un{mg/mL}$, which corresponds to $1\un{mM}$
of intrinsic DNA counterions~\cite{note2}. This value is also in accord with
results obtained by UV spectrophotometry on solutions of varying DNA and added
salt concentrations in that they gave a limit on the order of $1\un{mM}$ of
total counterions (intrinsic and added salt) below which double stranded DNA
denatures. We detected no change in the scaling behavior of fundamental lengths
of the two dielectric modes as we cross these solution conditions; nevertheless
a decrease in magnitude of the LF length scale is observed in the presence of
added salt in the vicinity of the denaturation threshold.

\begin{figure}
\onefigure[width=1.00\linewidth]{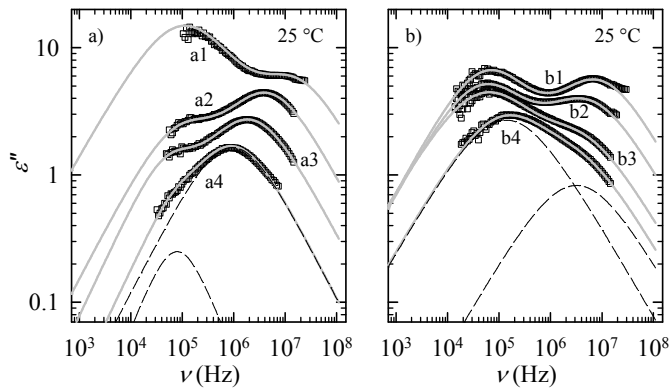}
\caption{Double logarithmic plot of the frequency dependence of the imaginary
part of the dielectric function ($\varepsilon^{\prime \prime}$) at $T=25\un{^\circ C}$
of a) pure water $146\un{bp}$ DNA solutions (for DNA concentrations a1-a4
(5, 0.5, 0.15, $0.05\un{mg/mL}$)) and b) $146\un{bp}$ DNA solutions with added
salt $I_{\mathrm{s}} = 1\un{mM}$ (for DNA concentrations b1-b4
(1.5, 0.8, 0.4, $0.3\un{mg/mL}$)). The full lines are fits to the sum of two
Cole-Cole forms (see text); the dashed lines represent the single form.}
\label{fig1}
\end{figure}

Fig.~\ref{fig1} shows the frequency dependent imaginary part of the dielectric
function for solutions with selected DNA concentrations. The results for pure
water short-fragment DNA solutions are shown in panel a), while results for DNA
solutions with added salt of ionic strength $I_{\mathrm{s}} = 1\un{mM}$ are shown in panel b).
The observed dielectric response is complex~\cite{Bordi} and the data can only
be successfully fitted to a formula representing the sum of two Cole-Cole forms
($\varepsilon(\omega) - \varepsilon_{\mathrm{HF}} = (\varepsilon_0 - \varepsilon_{\mathrm{HF}})/(1+(i\omega \tau_{0})^{1-\alpha})$),
equivalent to the Havriliak-Negami type with skewness parameter
$\beta=1$~\cite{Havriliak66}. The spectra consist of two broad modes that show a
symmetrical broadening of the relaxation time distribution function described by
the parameter $1-\alpha \approx 0.8$. The mode in the high-frequency region
(HF mode) has a strength $2<\Delta\varepsilon_{\mathrm{HF}}<15$ and is centered
in the range $0.3\un{MHz} < \nu_{\mathrm{HF}}< 15\un{MHz}$. The mode in the
low-frequency region (LF mode) has a strength $0.5<\Delta\varepsilon_{\mathrm{LF}}<50$
and for pure water solutions does not move much in frequency remaining centered
around $80\un{kHz}$ ($60\un{kHz} < \nu_{\mathrm{LF}} < 110\un{kHz}$). 

The polarization response of DNA solutions in the kHz-MHz range is due to an
oscillating flow of net charge associated with DNA counterions induced by an
applied ac field. Since the counterion displacement is facilitated through
diffusion, the dielectric response is basically characterized by the mean
relaxation time $\tau_0 \propto L^2/D_{\mathrm{in}}$, where $L$ is the
associated length scale, and $D_{\mathrm{in}}$ is the diffusion constant
of counterions. This constant is sufficiently well approximated by the diffusion
constant of bulk ions~\cite{Bordi,Wong} leading to a value of
$D_{\mathrm{in}}=1.33 \cdot 10^{-9}\un{m^2/s}$. In other words, the fit to
the Cole-Cole functions allows us to extract the characteristic time $\tau_0$
and calculate the corresponding length scale for each of the relaxation modes.
Similarly complex spectra have also been observed for long polydisperse DNA
semidilute solutions~\cite{Tomic06,Tomic07}. However the DNA concentration and
added salt dependence measured for short-fragment DNA are rather distinct,
indicating that mechanisms of counterion relaxation for short-fragment as
opposed to long Na-DNA solutions are not identical.

\begin{figure}
\onefigure[scale=0.70]{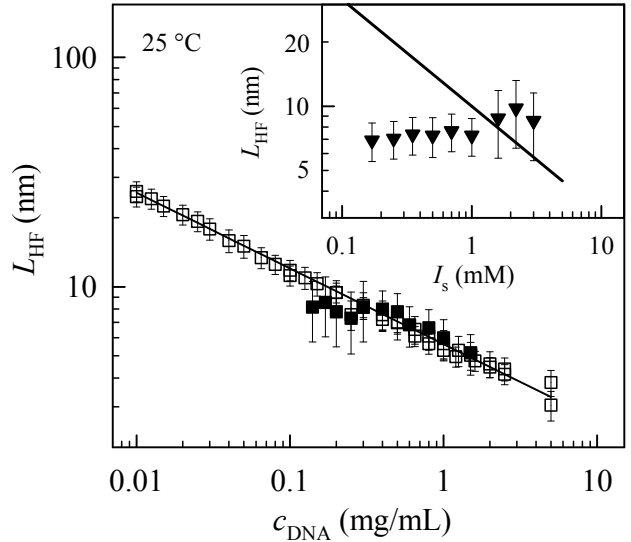}
\caption{Main panel: Characteristic length of the HF mode ($L_{\mathrm{HF}}$)
for pure water short-fragment DNA solutions (open squares) and for
short-fragment DNA solutions with added salt $I_{\mathrm{s}} = 1\un{mM}$ (full squares) as
a function of DNA concentration ($c_{\mathrm{DNA}}$). The full line is a fit to
the power law $L_{\mathrm{HF}} \propto c_{\mathrm{DNA}}^{-0.33}$.
Inset: $L_{\mathrm{HF}}$ for short-fragment DNA solutions \vs{} added salt
($I_{\mathrm{s}}$) for $c_{\mathrm{DNA}}=0.5\un{mg/mL}$ (full triangles). The full line
denotes Debye screening length for the investigated range of added salt.}
\label{fig2}
\end{figure}

Let us first describe the characteristics of the HF mode. For
pure water short-fragment DNA solutions the characteristic length $L_{\mathrm{HF}}$
increases with decreasing DNA concentration in almost three decades wide
concentration range (main panel of fig.~\ref{fig2}) following the power
law $L_{\mathrm{HF}} \propto c_{\mathrm{DNA}}^{-0.33}$ as a function of the DNA
concentration. In dilute solutions this scaling form is typical for the average
distance between chains~\cite{Dobrynin}. This result also confirms our claim in
the case of long, genomic DNA that the HF relaxation process describes the
collective structural properties of solution composed of many
chains~\cite{Tomic06,Tomic07}. It is noteworthy that although the DNA
double-helix appears to be denatured for $c_{\mathrm{DNA}} < 0.4\un{mg/mL}$, the
overall change in the prefactor of the scaling law is small and is within the
error bar of the experiment. This result is not too surprising. First, the same
scaling law is expected to be valid also for single stranded DNA, and second,
such a solution should contain twice the number of chains corresponding to only
about 20\% decrease in the average distance between chains, which is at the
resolution limit of our measurement. Also, the two chains that partake in the
organization of the common counterion cloud should probably remain in relatively
close proximity even after they are nominally dissociated. Furthermore even at
high DNA concentrations there is still no sign of a dilute-semidilute crossover,
confirming our estimate that $c^\ast$ is on the order of $1\un{mg/mL}$. Finally,
we remark that in semidilute solutions, but only in semidilute solutions, this
scaling form would be typical for charged chains with partially exposed
hydrophobic cores~\cite{Dobrynin}.

With added $1\un{mM}$ salt, the behavior of $L_{\mathrm{HF}}$
remains unchanged (main panel of fig.~\ref{fig2}), thus
$L_{\mathrm{HF}} \propto c_{\mathrm{DNA}}^{-0.33}$, as long as the concentration
of intrinsic counterions $c_{\mathrm{in}}$ (proportional to $c_{\mathrm{DNA}}$) is larger
than the concentration of added salt ions. At lower DNA concentrations, the
$L_{\mathrm{HF}}$ apparently shows a leveling off, with a limiting value close
to the Debye length appropriate for this salt concentration. A set of additional
data (inset of fig.~\ref{fig2}) for $c_{\mathrm{DNA}}=0.5\un{mg/mL}$ with
varying added salt concentration show that $L_{\mathrm{HF}}$ does not change
with $I_{\mathrm{s}}$ in most of the measured range. However, this behavior seems to remain
even in the limit when added salt concentration is larger than the concentration
of DNA intrinsic counterions, showing $L_{\mathrm{HF}}$ values apparently above
the corresponding Debye length. This is in contrast to $1\un{mM}$ data (shown
in the main panel of fig.~\ref{fig2}). We are tempted to believe that this
apparent contradiction is related to a poorer accuracy of these data, in
comparison to $1\un{mM}$ ones, due to the progressive merging of the HF and LF
modes when one approaches the regime of high added salt.

\begin{figure}
\onefigure[scale=0.70]{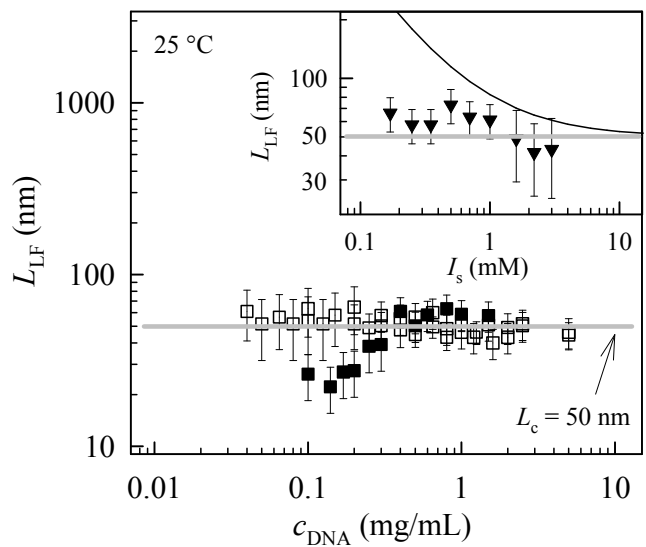}
\caption{Main panel: Characteristic length of the LF mode ($L_{\mathrm{LF}}$)
for pure water short-fragment DNA solutions (open squares) and for
short-fragment DNA solutions with added salt $I_{\mathrm{s}} = 1\un{mM}$ (full squares) as
a function of DNA concentration ($c_{\mathrm{DNA}}$). The full line denotes the
contour length $L_{\mathrm{c}} \approx 500\un{\mbox{\AA}}$.
Inset: $L_{\mathrm{LF}}$ for short-fragment DNA solutions with
varying ionic strength of added salt for $c_{\mathrm{DNA}} = 0.5\un{mg/mL}$
(full triangles). The grey and full lines denote the contour length of studied
short-fragment DNA and the persistence length as predicted by the OSF theory,
respectively (ref.~\cite{Odijk77}).}
\label{fig3}
\end{figure}

Second, we address the LF mode. For pure water short-fragment DNA solutions, the
characteristic length $L_{\mathrm{LF}}$ remains approximately constant in more
than two decades wide concentration range (open squares in fig.~\ref{fig3}) at
the level of $L_{\mathrm{c}} \approx 500\un{\mbox{\AA}}$. The latter value suggests that in
this regime $L_{\mathrm{LF}}$ is proportional to the contour length of the
polyelectrolyte chain~\cite{note3}. This result confirms that the LF relaxation
represents a single chain property~\cite{Tomic06,Tomic07}. For DNA solutions
with added salt $I_{\mathrm{s}} = 1\un{mM}$ (full squares in fig.~\ref{fig3})
$L_{\mathrm{LF}}$ is the same as for pure water DNA solutions at high DNA
concentrations. At the concentrations of intrinsic counterions $c_{\mathrm{in}}$
(proportional to $c_{\mathrm{DNA}}$) smaller than the concentration of added
salt ions, $L_{\mathrm{LF}}$ starts to deviate from the
$L_{\mathrm{LF}} \propto L_{\mathrm{c}}$ behavior and decreases to attain a value of about
$250\un{\mbox{\AA}}$. Additional data (inset of fig.~\ref{fig3}) for
$c_{\mathrm{DNA}} = 0.5\un{mg/mL}$ with varying added salt show that
$L_{\mathrm{HF}}$ does not vary with $I_{\mathrm{s}}$ in most of the measured range of
added salt. When added salt concentration is larger than DNA concentration, the
behavior of $L_{\mathrm{LF}}$ indicates only a minor decrease, in contrast to a
more substantial one shown by $1\un{mM}$ data (main panel of fig.~\ref{fig2}).
This discrepancy might be ascribed, similarly as for the HF process, to a poor
accuracy of former data, which are severely influenced by high salt environment.
The added-salt-independent behavior in the limit of low added salt is not
surprising since the contour length of the short-fragment DNA chains is on the
order of the DNA intrinsic persistence length. This fact immediately excludes
the effects of electrostatic interactions on the persistence length as predicted
by OSF theory~\cite{Odijk77} and shown for comparison in inset of fig.~\ref{fig3}.

\begin{figure}
\onefigure[scale=0.70]{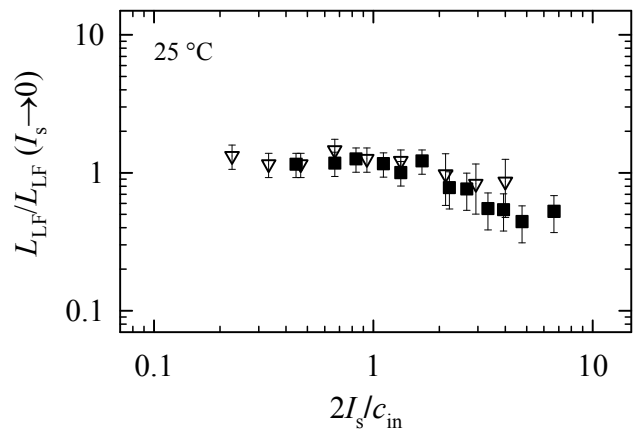}
\caption{Characteristic length of the LF mode ($L_{\mathrm{LF}}$) normalized
with the value in pure water solutions $L_{\mathrm{LF}}(I_{\mathrm{s}} \to 0) \approx L_{\mathrm{c}}$
\vs{} added salt concentration normalized by the concentration of intrinsic
counterions ($2I_{\mathrm{s}}/c_{\mathrm{in}}$). Data are for one representative DNA concentration
$c_{\mathrm{DNA}}=0.5\un{mg/mL}$ and varying $I_{\mathrm{s}}$ (open triangles) and for
$I_{\mathrm{s}} = 1\un{mM}$ and varying DNA concentrations (full squares).}
\label{fig4}
\end{figure}

In the dilute regime it is expected that the average distance between chains
($R_{\mathrm{cell}}$) is larger than the contour length. Our data
(see fig.~\ref{fig2} and \ref{fig3}, main panels) are at variance with
this expectation: although $L_{\mathrm{HF}}$ scales as expected for the average
distance between chains, it is smaller than $L_{\mathrm{LF}}$ in the whole range
of DNA concentrations. In order to rationalize this finding, we deem it
plausible that $L_{\mathrm{HF}}$ corresponds to a shorter scale inside the
spherical cell zone of the size $R_{\mathrm{cell}}$ around the polymer. Indeed,
theoretically the dilute regime is modelled by placing the polymer at the center
of a cell of size $R_{\mathrm{cell}}\propto c^{-0.33}$ that is subdivided into
two zones~\cite{Deshkovski,Dobrynin}: a smaller cylindrical one, inside which
the electrostatic interaction energy is large, and a larger spherical zone,
where the electrostatic interactions are described by the low coupling
Debye-H\"{u}ckel approximation. As a result, the response of DNA
counterions to an applied ac field would be mostly confined to a smaller
cylindrical volume within the large spherical volume of radius equal to
the average distance between chains. We remark that a similar relationship
between $L_{\mathrm{HF}}$ and the contour length in dilute regime, although not
discussed, can be discerned in the data reported by Ito \al{}\ (ref.~\cite{Ito90})
and by Katsumoto \al{}\ (ref.~\cite{Katsumoto07}).

It is noteworthy that the contour length remains the relevant single chain
length scale as long as the concentration of intrinsic counterions is larger
than the concentration of added salt, as can be clearly noticed from fig.~\ref{fig4}. 
In other words, the single chain length scale is independent of the DNA
concentration due to the fact that in the dilute regime interchain interactions
are negligible compared to intrachain interactions, in contrast with the
behavior of long DNA chains in semidilute solutions ~\cite{Tomic06,Tomic07}.
The data for $L_{\mathrm{LF}}$ deviate from the $L_{\mathrm{c}}$ only in the high salt
limit, for $2I_{\mathrm{s}} > 2 c_{\mathrm{in}}$, where we expect that the effects of added salt
become dominant. In this regime on addition of salt, (fig.~\ref{fig4}), we note
that the fundamental single chain length scale shrinks in size, becoming smaller
than the nominal contour length of the chain. Taking into account that the short
fragments of DNA are almost exactly one persistence length long, it is difficult
to envision any decrease in the rigidity, as quantified by the persistence
length, that would lead to a smaller effective contour length. A possible
speculative explanation for this strange behavior could be sought in the
incipient dynamic dissociation of the two strands of DNA that is fully
established for DNA concentrations below $0.4\un{mg/mL}$. Short bubbles of
separated strands along DNA and/or pronounced fraying at the two ends could
shorten its effective contour length which would then correspond to the measured
fundamental single chain length scale.

In summary, our results demonstrate that the LF and HF processes, detected by
dielectric spectroscopy measurements of short-fragment DNA solutions, are
associated with structural properties of single chains as well as collective
properties of the solution composed of many chains, respectively. In dilute
conditions and in the low added salt limit, the characteristic lengths of the
two relaxation modes are given either by the contour length of a single DNA
chain or the reduced average distance between chains of the whole solution,
consistent with the two zone model of counterion condensation~\cite{Deshkovski,Dobrynin}.
In the high added salt limit for $2I_{\mathrm{s}} > 2 c_{\mathrm{in}}$ our data
indicate that the added salt effects are relevant for both length scales. For
the single chain length scale, salt apparently facilitates the formation of
denaturation bubbles and/or fraying at the two ends close to DNA concentrations
where the strand separation is clearly seen. For the collective length scale, on
addition of salt the Debye screening length seems to take the role of the
fundamental length scale for the whole solution. The latter statement should be
taken with caution, however, due to poor accuracy of the data at higher added
salt. Finally, our results indicate that, at the dilute DNA conditions and in
low added salt regime, the same fundamental length scales are obtained, even
under the conditions where the dissociation of the DNA strands is likely to be
expected.

\acknowledgments
We would like to thank Don Rau and Adrian Parsegian for valuable and
illuminating discussions. This work was supported by the Croatian Ministry of
Science, Education and Sports under grant 035-0000000-2836. R.P.\ would like to
acknowledge the financial support of the Agency for Research and Development of
Sloveniaunder grant P1-0055(C) and partial financial support by the Intramural
Research Program of the NIH, National Institute of Child Health and Human
Development.


\begin{thebibliography}{0}
\bibitem{note1}
$c^\ast$ is given by the concentration where there is only one polymer molecule
in the volume of a polymer globule $c^\ast = \textrm{molecule mass}/V_{\mathrm{c}}$, where
molecule mass is $N \cdot m_{\mathrm{bp}}$, and
$V_{\mathrm{c}} \approx L_{\mathrm{c}}^3 = N^3 \cdot a^3$; $m_{\mathrm{bp}}$ is a mass of a base pair
$\approx 10^{-18}\un{mg}$, $N=146$, $L_{\mathrm{c}}=N \cdot a$; $a=3.4\un{\mbox{\AA}}$;
$a$ is the monomer size.
\bibitem{note2}
The concentration of intrinsic counterions is given by
$c_{\mathrm{in}} \left[\mathrm{mM}\right] = c_{\mathrm{DNA}} \left[\mathrm{mg/mL}\right] \cdot 3\un{\mu mol/mg}$.
\bibitem{note3}
For $c_{\mathrm{DNA}} < 0.2\un{mg/mL}$ the LF mode decreases in strength and
merges into the larger HF mode. In the range
$0.04\un{mg/mL} < c_{\mathrm{DNA}} < 0.2\un{mg/mL}$, two-modes fits with
constraint on the relaxation time of LF mode were done in order to keep
$L_{\mathrm{LF}}$ close to $L_{\mathrm{c}}$. The constraint did not influence the
parameters obtained for the HF mode. For $c_{\mathrm{DNA}} < 0.04\un{mg/mL}$,
the LF mode is of negligible strength and completely overwhelmed by the HF one,
so that only single mode fits to the HF mode were performed.
\end{thebibliography}
\end{document}